# Benefits of AWS in Modern Cloud

Sourav Mukherjee

Senior Database Administrator &

PhD student at University of the Cumberlands

Chicago, United States

**Abstract**

This article gives an overview of the benefits of AWS in the modern cloud. Cloud computing is performing well in today's World and boosting the ability to use the internet more than ever. Cloud computing gradually developed a method to use the benefits of it in most of the organizations. It is very demanding in all businesses tasked with improving the quality of service reducing costs as the organization pays for the service only what they consume based on the incoming and outgoing traffic.

**Introduction:**

The benefits of AWS in the modern cloud are huge. Data protection, regulatory compliance, quantifiability, flexibility, cost-effectiveness, multiple storages, auto-scaling, access to the data anytime, data-centric encryption, high-performance processing are few benefits of AWS cloud. Let's understand the importance of the AWS cloud in detail to get the best idea of it.

**Data Protection:** Data is the most important asset in any organization. Data leakage can cause a huge loss in the organization, so every organization thinks about data privacy at the first point to protect their sensitive data. Not following the guidelines of data protection may cause loss or theft of company intellectual property, damage to the organization's reputation, corporate or individual penalties and compromising the system to hacking or malware infection vulnerabilities. The authorized use of cloud environment in the organization and the capability to transfer sensitive evidence into and throughout, the cloud plays a vital role for industries to function and work proficiently, speedily and without any restrictions. But this capability must be maintained by an inclusive data safety approach which AWS cloud maintains by using security controls and processes. The below AWS best approaches can be followed for data storage and protection:

- Implement data encryption/hashing on the device and server.

- Sensitive local data stored encrypted with user secret that encrypts the data encryption key.
- Use NIST (National Institute of Standards and Technology) approved encryption standard algorithms to encrypt the sensitive data.
- Encryption keys shall never be in RAM. Instead, keys should be generated real-time for encryption/decryption as needed and discarded each time.
- No sensitive data (e.g. passwords, keys etc.) in cache or logs.
- Use remote wipe APIs.
- Do not reveal UDID (unique device identifier), MSISDN (Mobile Station International Subscriber Directory Number), IMEI (International Mobile Equipment Identity) and PII (Personally Identifiable Information).

**Access Management:**

- **Regulate access to cloud resources at a granulated level:**

  Using enlightened strategy users and groups can be created to regulate the access management, for example, user, resource, IP address, time of day that means the deployment is done using AWS cloud is secure

  AWS IAM permits to generate and accomplish users and groups, as well as use authorizations to regulate access to AWS resources like Amazon S3 storage stacks, Amazon EBS snapshots, or Amazon DynamoDB tables.

- **Integrate with current individuality and access management systems:**

  Combining with the current individuality and access management systems means that there is no need to go through the procedure of generating equivalent sets of individualities in the cloud. Individualities in the current systems can be used to give access to the resources in the AWS cloud.

**Regulatory Compliance:** Compliance needs to be established for every day checkpoints, not just during inspections and audits. Most of the organizations are accomplishing compliance by going back to security essentials, by automating the processes for artificial intelligence, cleaning the data into their operations. The below approaches for Authentication, Authorization and Session Management can be followed to prove the compliance and minimize the business risks.

- Strong password policy
- Validate password and sessions if the application needs to work in offline mode
- Use salted password
- CAPTCHA during registration
- Unique session tokens to form valid and unique message payloads
- Corporate approved encryption/hashing algorithms
- Two-factor authentications (in case of financial transactions to be performed.)

- Lower timeout for the inactive session
- Server-side authentication for sensitive transactions.
- Validate all messages/payloads received at the backend / mobile application server and prevent message replay attacks. These messages/payloads should be encrypted and should have a combination of padding elements, session identifiers, and timestamps.

**Flexibility:** The key benefits of cloud computing are its flexibility. Business in the organization can scale up or scale down and the data loads may need quick modification which is very flexible in AWS cloud. That way cloud computing permits the employees to be more flexible. Employees can access files anywhere using web-enabled devices such as laptops, smartphones, notebooks etc. The capability to instantaneously share documents and other files over the internet can also assist support in the association. Cloud computing allows the use of mobile technology. Enterprise mobility management tools can provide valuable administrative capabilities and protect the organization from phone loss, accidental data loss or weak passwords, necessary visibility into today's modern security risks, including malware and other device-centric attacks. AWS cloud permits the business to straightforwardly upscale or downscale its current resources to accommodate business requirements which enables to support the business development without exclusive changes of current systems. Flexibility is one of the key factors why companies move their business to the cloud.

**Cost Effectiveness:** One of the most important benefits of cloud computing is considerable savings in organizations cost. By moving to cloud computing, industries can save extensive capital costs through minimizing spending on infrastructure, equipment, and software. Cloud permits to rent extra processing power over the Internet without having to use million-dollar machines as servers. Spending a lot of money on the hardware, software or licensing and renewal fees, many companies are showing interest to move to cloud environments and getting benefited saving the cost. The company just needs to pay on the usability and the traffic, if the platform is not used then the company saves the money. AWS cloud needs minimum asset the start the service and less expensive than the on-premise installations.

**Secure backend services and platform:** Secure backend services and platform: The benefit of the AWS cloud is that it permits customers to scale and transform while sustaining a protected environment. Customers pay only for the facilities they use, meaning that customer has a secure backend service and the platform, but deprived of the upfront expenditures, and at a lower cost than in an on-premises environment. There are few AWS cloud benefits for secure backend services listed below.

- Implement Protected Backend API'S or facilities
- Secure data allocation between the cloud and web-server back- ends and other external interfaces
- Server and infrastructure inurement
- Maintain and monitor application server logs
- Access control for cloud platform

**Increase productivity:** Previously the time is mostly spent on the software installation, working on the maintenance of the product and take the back up on daily basis. Cloud has solved most of these problems where software installation is not needed, most of the maintenance is done by Amazon team, back up are automated. Anyone from anywhere with proper access can log in to the company's cloud platform. A lot of time can be solved using a cloud platform which increases productivity. The acceptance of the cloud has been determined by the digitization of the Corporate World, which has exponentially added to the amount of data, plans, and arrangements that an organization needs to manage to keep up. The cloud proposes the best way to keep the business planned and ground-breaking, and productions have stated back strong outcomes. A survey was initiated and found that 79% of the users reported higher revenue growth using a cloud platform.

**Increase Scalability:** To know how the organization will grow and what is the future is one of the business's major challenges. The cloud has opened prospects for the organization to grow in their business. The cloud is scalable which is essential for a business to grow. Whether more resources are needed or less, cloud instances can promptly adapt the needs. On-premise infrastructure takes days or weeks to set up the connection and maintain the hardware and software. Comparatively, the cloud is very easy as this resides over the internet, so it increases the scalability. AWS also supports Dynamo DB auto-scaling; the capacity allocation can be optimized for cost and usage. Read/Write capacity units for each of the DynamoDB tables can be consumed and analyzed to determine the minimum, maximum capacity allocation for autoscaling. For different backbiting and data prep, scalability is a major factor, it's mandatory to tune the system for maximum performance and scalability. There are usually a few key considerations when it comes to scalability for a performant environment to support enterprise data backbiting which AWS cloud supports:

- When there is huge volume of data
- Data access interface (e.g. Files, APIs)
- Number of concurrent users
- Type of backbiting operations and use cases (e.g. structuring, blending, profiling)
- Hardware configuration (number of nodes, network setup)

**No to Piracy:** Using AWS cloud platform can prevent software piracy incessantly. The illegal copying of the data is a violation of any company. It may cause loss or theft of a company's intellectual property and damage the company reputation. So, every company needs to follow some security practices to reduce the risks and mitigate threads. New cloud environments are treated establishing new data centers. AWS cloud has all applicable security controls and processes in place which solves the problems of security risks and mitigate the threads.

**Advanced technology and career opportunities:** Cloud is the presence of IT. AWS Cloud allows organizations to arrange advanced real-time services and accomplish significant profits and productivity developments. Understanding of AWS cloud is not that difficult as Amazon documentation is available easily. Organizations need a great arrangement of additional control and flexibility over the technologies they develop, they needed the AWS cloud platform that could provide more flexibility also innovative for the future. The cloud platform is scalable, extensible, and scattered architecture which is easy to build and maintain. Most of the maintenance is done by the Amazon team and back up are automated. Manual tasks can be reduced which is a progressive approach in all organization now a day. It is not overloaded the network and supports strong security management, system validation, and easy to get subscription authorization as cost is based on the usages. Anyone from anywhere with proper access can use the online cloud. As the cloud is over the internet, employees needn't install and run on the computer as on-premise services. The career opportunities are exiting as there are so much to learn, innovate in this platform.

**Make life easier:** Life is full of taking various decisions. If you run a business, then you must contribute a lot on daily basis to succeed in the business. Same for any organizing as every organization faces challenges in the business and taking the right decision is very important. With the benefit of AWS cloud, anyone can be confident that he is taking the very right decision for the company growth and to achieve the goal. The below points will explain why AWS cloud make life easier.

- **Financial Savings:** In on-premise software company must own the hardware and servers which are expensive for long run and maintenance is tough. But in the cloud, the payment is made based on the usages of the services. Think about if a company could reduce the expenses by preventive their requirement for the hardware they use, IT system maintenance, downtime, and even the quantity of energy usage. Using the AWS cloud, over a period, it can offer some incredible saving assistance. Cloud permits to rent extra processing power over the Internet without having to use million-dollar machines as servers.

- **Flexibility:** As mentioned above flexibility of cloud is one of the reasons why it makes life easier. To work efficiently business needs to familiarize most circumstances in a timely manner to succeed in the business. Anyone with proper access can log in to cloud from anywhere anytime and move the information one environment to another environment – no longer controlled by restrictions of using hardware.

- **Disaster recovery:** With the benefit of a cloud-based platform, the company is secured as data is protected and to succeed in the industry is easy to compare to using the on-premise platform. AWS cloud assists to resolve the issues encountered very speedily and proficiently so the customer needs to take less tension for troubleshooting any issue. If the electricity has gone and data is not saved, they must start from scratch. Using the AWS cloud, the company no longer needs to concern losing the data when not saved. Once the data is uploaded in a cloud it is protected in a secure location which makes life easy in any business. By using AWS cloud, a company makes multifaceted and time-consuming adversity retrieval plans of the past.

- **Communication and Association:** Communication and association between employees are essential for company growth. One incredible advantage of cloud computing is the capability to progress communication and collaboration within a group. By adapting to the cloud platform, employees have all the required resources they need to work proficiently from anywhere irrespective of where they are positioned or which strategies they're using. The cloud will permit instant circulation of pertinent material to everybody within the organization who have access to the loud platform, guaranteeing continuous output.

- **Multiple Storages:** Does anyone remember about the floppy disk as a storage device? The DVD and 2 GB of the hard disk was a big thing during the starting of Y2K. Now a day a normal system uses more than 1 terra byte of data. Very few of us use a pen drive to copy any file from one computer to another. Cloud is the present and cloud is the future. Today no one needs to worry about storing the data as the cloud is easily accessible. Even in smartphones, data can be stored in the cloud so if the phone is lost, the data is still available in the cloud and accessible from anywhere with the cloud user id and password. Also, needn't be worried about data security.

**Storage options in AWS cloud:** Designers of traditional, on-premises IT infrastructures and applications have many possible data storage selections, including the following:

- **Memory** - In-memory storage, such as a file, object, databases caches, and RAM disks, deliver quick access to the data.

- **Message Queues** - Provisional robust storage for data not sent synchronously between computer systems or application components.

- **Storage area network (SAN)** - NAS storage offers a file-level boundary to storage that can be collected across numerous structures. NAS tends to be slower than either SAN or DAS.

- **Direct-attached storage (DAS)** - Local hard disk drives or collections of data exist in each server deliver advanced performance than a SAN, but lower robustness for provisional and tenacious files, database storage, and operating system (OS) boot storage than a SAN.

- **Network attached storage (NAS)** - NAS storage offers a file-level boundary to storage that can be collected across numerous structures. NAS tends to be slower than either SAN or DAS.

- **Databases** - Organized data is classically stored in a database, such as a traditional SQL relational database, a NoSQL non-relational database, or a data warehouse. The fundamental database storage naturally exists in SAN or DAS devices, or in some cases in memory.

- **Backup and Archive** - Data booked for backup and archival purposes is naturally stored on non-disk media such as tapes or visual media, which are generally stored off-site in isolated protected sites for disaster recovery.

All the above storage options vary in performance, robustness, and price, as well as in their interfaces. Architects must study and understand all these aspects when recognizing the right storage resolution for the task at hand. When it comes to AWS cloud, they offer multiple storage options Each has an exclusive combination of performance, robustness, accessibility, price, and interface, as well as other features such as scalability, flexibility, elasticity. These extra features are crucial for web-scale cloud-based platforms.

- **Amazon Glacier:** Amazon Glacier is for long-term storage which is very secure and durable object storage. 10 GB of Amazon Glacier data recoveries per month for free. The free layer payment can be used at any time during the month and applies to standard recoveries of the stored files.

- **Amazon S3:** Amazon Simple Cloud Storage Service is to store the standard files which are secure, durable and scalable object storage infrastructure. Mostly one-month recent data can be stored in S3, other old files move to Amazon Glacier for future retrieval. S3 can have
  - 5 GB of standard storage
  - 20,000 Get requests
  - 2,000 Put requests

- **Amazon Elastic Block Storage (EBS):** Amazon Elastic Block Storage can be insistent, durable, low-latency block-level storage volumes for EC2 instances. Elastic Block Storage can have
  - 30 GB of Amazon EBS: any combination of general purpose (SSD) or magnetic
  - 2,000,000 I/O (with EBS magnetic)
  - 1 GB of snapshot storage

**Amazon EC2 Instance Storage:** Temporary block storage volumes for Amazon EC2 (Elastic Compute Cloud) virtual machines. EC2 instance storage offers temporary block-level storage for the instances and these storages are placed in the disk that is physically attached to the host computer.
StorReduce supports storing unstructured data to Amazon Simple Storage Service (Amazon S3) or Amazon Glacier on AWS to minimize the cost. The below diagram shows how the

StorReduce works.

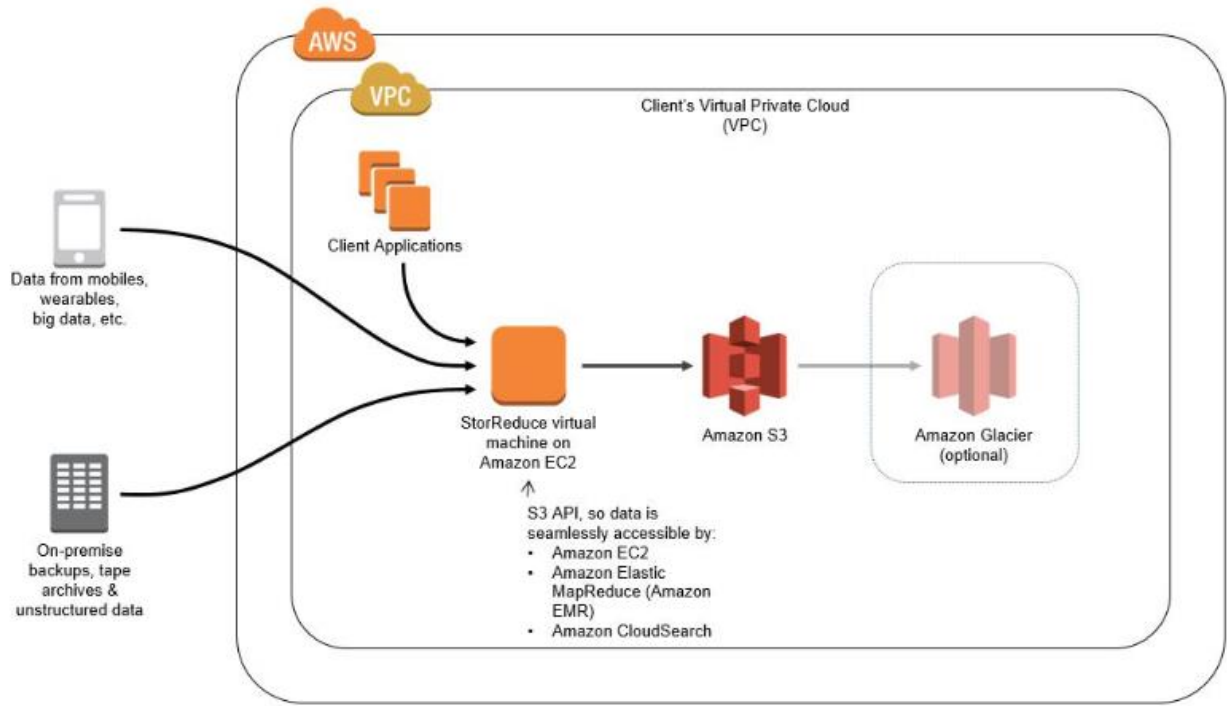

Fig1: Workflow of StorReduce [14]

**AWS Import/Export:** This is a service which is used to transfer a large volume of data from the physical storage device into AWS.

**AWS Storage Gateway:** Permits on-premises environments to use AWS cloud storage.

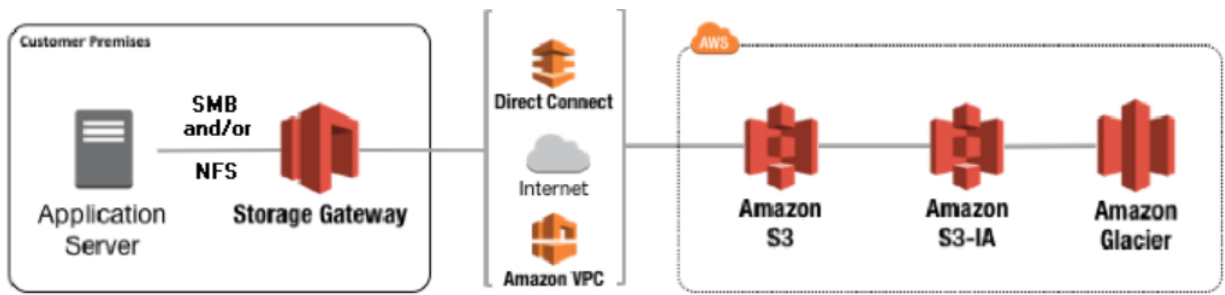

Fig2: Overview of file storage deployment for Storage Gateway [15]

**Amazon CloudFront:** Globally circulated network of proxy servers.

**Amazon SQS:** Simple Queue Service (SQS) is message queue service used by distributed systems to exchange messages.

**Amazon RDS:** Relational database service for MySQL, Oracle, MS SQL server.

**Amazon DynamoDB:** NoSQL database which is highly predictable and scalable.

**Amazon ElastiCache:** This is a fully managed in-memory data store and cache service.

**Amazon Redshift:** Data warehouse service fully managed, fast and powerful.

**Databases on Amazon EC2:** Self-managed database on an Amazon EC2 instance.

**Backup Options:** AWS backup is a complete backup service that makes it easy to integrate and automate the back up of data across AWS cloud platform using AWS storage gateway. AWS back up completely managed by policy back-up solution, simple to use, allow to meet the compliance requirements in the business and very secure. There are huge benefits, some are listed below.

- **Centrally manage backups:** Arrange backup strategies from a central backup console, streamlining backup management and creating it easy to confirm that the application data across the AWS platform is backed up and secured. AWS Backup's central console can be used, APIs, or command line interface to backup, reinstate, and set backup maintenance policies across AWS platform in the cloud and on-premises using the AWS Storage Gateway.

- **Automate backup processes:** Time and money can be saved, and manual error can be avoided using the automate backup processes. AWS Backup's completely manageable, policy-based solution. AWS Backup offers automated backup agendas, maintenance, and development, eliminating the need for routine scripts and manual procedures. With AWS Backup, backup policies can be applied to the AWS resources by simply classifying them, making it easy to develop the backup strategy across all AWS resources and confirming that all application data is properly backed up.

- **Improve backup compliance:** By applying the backup policies, encode the backups, and review backup actions from a centralized console to assists meet up the backup compliance necessities. Backup policies make it simple to line up the backup strategy with internal or monitoring necessities. AWS Backup secures backups by encrypting data during transfer and at rest. Joined backup action logs across AWS services makes it easier to achieve compliance audits. AWS Backup is PCI and ISO compliant as well as HIPAA eligible.

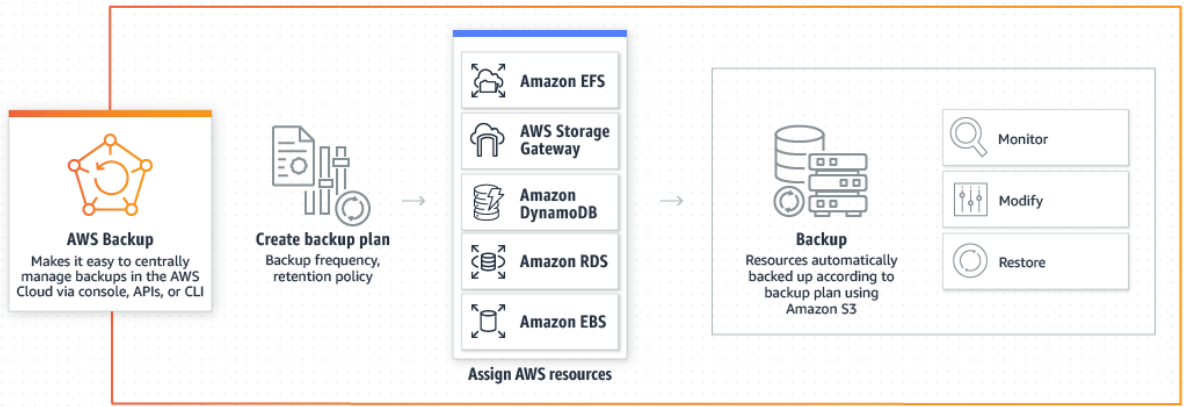

Fig3: Workflow of automatic backup in AWS cloud [17]

**Speed up the workflows:** Speed up the workflows: Another benefit of cloud that is it helps to take lesser time to complete the project. Sometimes the project takes a longer time to complete an employee become less productive. Cloud solves that problem as it takes lesser time, so employees can get motivated doing the work. With robust teamwork, project supervision, client satisfaction, and using other cloud tools, employees can complete the tasks which take less time. Anyone with proper access can log in to the cloud environment from anywhere which speed up the workflows. Employees can work smartly to get the work done. If there is any error, the troubleshooting is very easy as the clear logs will be CloudWatch. If needed Amazon support can be involved to troubleshoot the issue so overall, it's easy to maintain which move faster in any project using cloud platform.

A business procedure is characterized as a workflow shown in the diagram below. Applications frequently include a workflow as stages that must take place in a predefined direction, with prospects to correct the flow of information based on certain results or special suitcases.

.

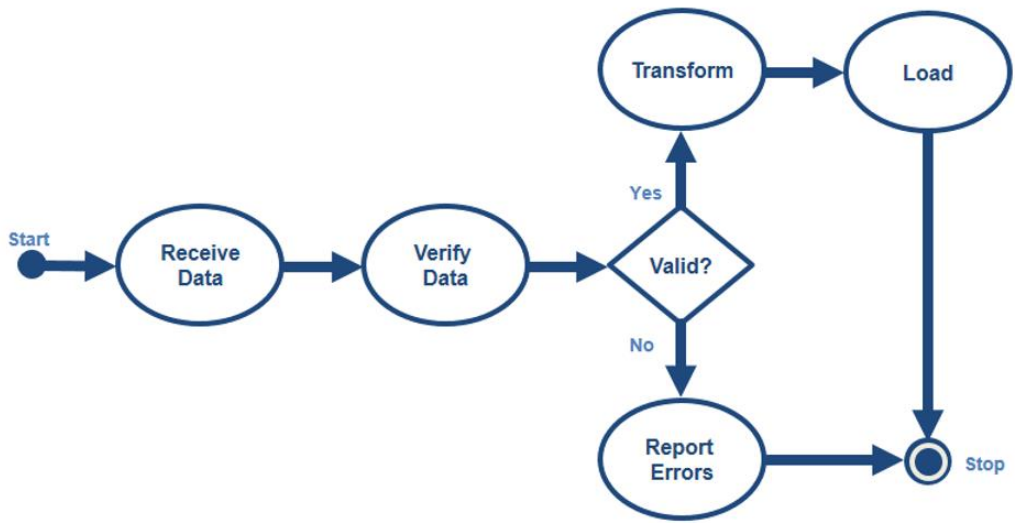

Fig4: ETL workflow [18]

**Minimize IT workload:** Using cloud platform the IT team in the company no longer have to be busy with the IT infrastructure, servers or computers, installing new software licenses as everything will be in the cloud. A central management system is introduced in the cloud which lets immediate pushes for any updates and licensing, and the Amazon team has the ability to fix the problems remotely which minimize the IT workload in the organization.

- Cloud has the capability to schedule automatic jobs and application deployment which reduces the manual work.
- Understand the potential of big data and accomplish workloads in accessible ways.
- Secure development for file transfer operations, immediate status visibility, and automatic retrieval of data.
- Allow DevOps association with a Jobs-as-Code method for quicker submission and deployment series periods.
- No downtime upgrades to remove business disruption and risk.

Let's take an example of how AWS cloud minimizes the workload. An important part of any web application is static content. This contains tapes, image, text, and other content that varies occasionally, web servers are migrated to EC2 instances and host all contents static and dynamic. Introducing static content from an EC2 instance experiences several costs including the instance, EBS volumes, and likely, a load balancer. By moving static content to S3, the cost can be expressively reduced the amount of computing required to host your web applications. In many cases, this change is non-disruptive and can be done at the DNS or CDN layer, needing no change to your application.

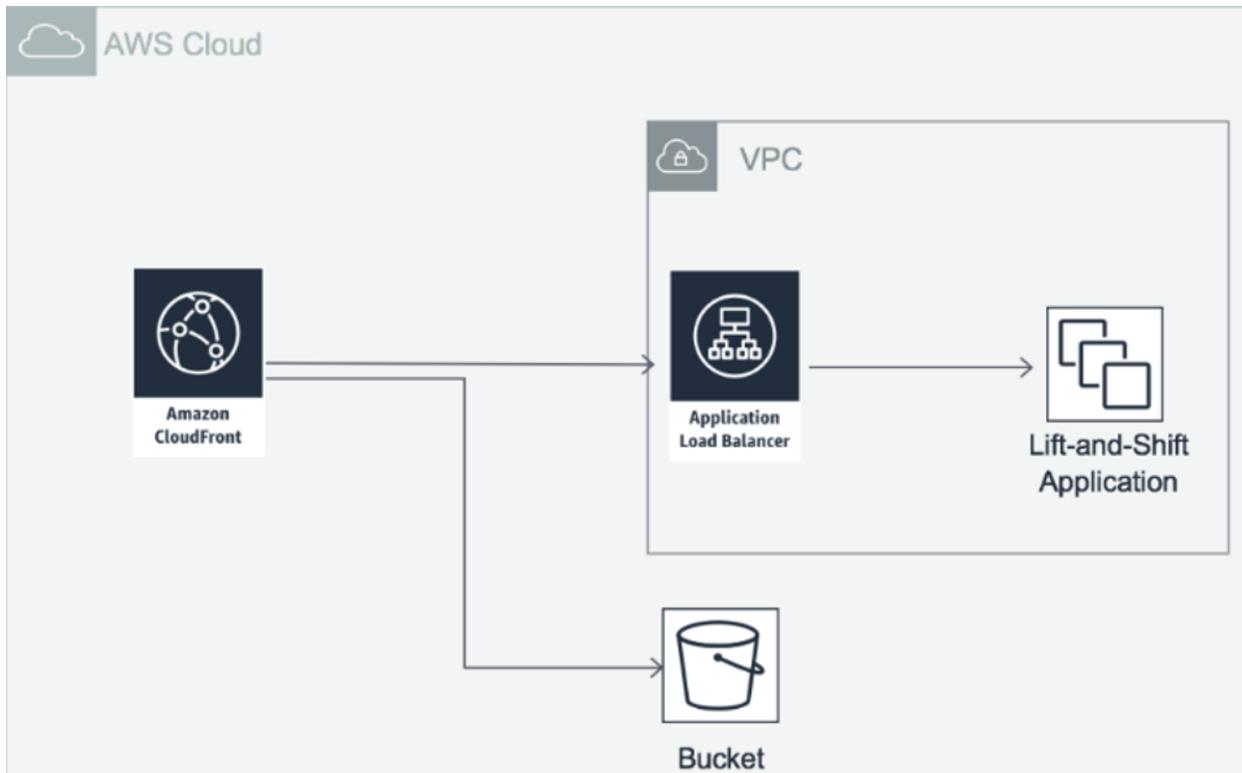

Fig5: Reducing web hosting costs with S3 static web hosting [19]

**Different databases in AWS Stack:** AWS database stack gives access to the competences of a familiar MySQL, Oracle, SQL Server, or PostgreSQL database engine. Scripts, applications, and tools that already being used with the existing databases can be used with Amazon RDS. Automatically patches the AWS database software and backs up the company data. 7 database types are used under AWS now.

- Amazon Aurora - Relational Database Built for the Cloud
- Amazon RDS - Managed Relational Database Service for MySQL, Oracle, SQL Server, and PostgreSQL

- Amazon DynamoDB - Fast, Predictable, Highly-scalable NoSQL data store.

- Amazon ElastiCache - In-Memory Caching Service.

- Amazon Redshift - Fast, Powerful, Fully Managed, Petabyte-scale Data Warehouse Service.
- Amazon Neptune - Fast, Reliable, Fully-managed graph database service that makes it easy to build and run applications that work with highly connected datasets.

- AWS Database Migration Service – This service helps to migrate the databases to AWS effortlessly and securely, the source database stays fully functioning during the migration, reducing downtime to applications using the database.

**Amazon RDS (Relational Database Service):**

- Easy to setup and manage & Scale a relational DB to Cloud.
- Cost-effective and resizable capability. Compensate only for the resource you use.
- It can automate time-exhausting administration duties.
- Users can focus on applications rather focusing on true admin tasks.
- Provides fast performance, high availability, security and compatibility they need.
- Easier to go from project conception to deployment.
- Scale your database's compute and storage resources with just a mouse click.
- Availability. Can run on the same highly reliable infrastructure used by other Amazon Web Services.
- Fast. Offers performance on parity with commercial databases at 1/10th the price.
- Secure. The RDS process creates it simpler to control network access to your database.

**Amazon DynamoDB:**

- It is a Fast and flexible NoSQL database service that offers consistent, single-digit millisecond latency at every scale.
- Supports both document and key-value store models.
- Flexible data model and reliable performance. Automatic scaling of data capacity makes it a great fit for mobile, web, gaming, ad tech, IoT, and many other applications.
- Fully Managed. No longer necessary to be concerned about database management duties.
- Fine-grained Access Control. Combines with AWS Identity and Access Management (IAM) for easy access control.
- Event Driven Programming. Integrates with AWS Lambda to provide Triggers which enables to architect applications that automatically react to data changes.
- Highly Scalable. Spontaneously scales the facility up or down, as product request volumes go up or down.
- DynamoDB Accelerator or DAX is a completely controlled, highly accessible, in-memory collection. Can reduce response times from milliseconds to microseconds, even at millions of requests per second.

**Amazon Aurora:**

- It is a combination of MySQL and PostgreSQL compatible relational database developed for the cloud.
- Offers 5X the output of standard MySQL and 3X the output of standard PostgreSQL operating on the same hardware.

- Fault-tolerant, Distributed, and self-healing storage capacity that auto-scales up to 64TB per database instance.
- Delivers high performance and availability with up to 15 low-latency read replicas, point-in-time recovery, continuous backup to Amazon S3, and replication across three Availability Zones.
- Offers greater than 99.99% availability.
- Provides multiple levels of security for your database. These include network isolation using Amazon VPC.
- It is fully managed by Amazon Relational Database Service (RDS). You do not have to worry about database management duties such as hardware provisioning, software patching, setup, configuration, or backups.
- MySQL and PostgreSQL compatibility make Amazon Aurora a compelling target for database migrations to the cloud.

**Amazon ElastiCache:**

- Offers fully managed Redis and Memcached. EXTREME PERFORMANCE - It works as an in-memory data store and cache to support the most demanding applications requiring sub-millisecond response times.
- Seamlessly deploy, operate, and scale popular open source compatible in-memory data stores.
- Build data-intensive apps or improve the performance of your existing apps by retrieving data from high throughput and low latency in-memory data stores.
- The popular choice for Gaming, Ad-Tech, Financial Services, Healthcare, and IoT apps.
- Fully managed. You do not need to perform management duties such as hardware provisioning, software patching, setup, configuration, monitoring, failure recovery, and backups. It always monitors your clusters to keep your workloads up and running so that you can focus on higher value application development.
- SCALABLE. It can scale-out, scale-in, and scale-up to meet fluctuating application demands. Write and memory scaling is supported with shading. Replicas provide read scaling.
- Amazon Elastic Cloud is based on REDIS.

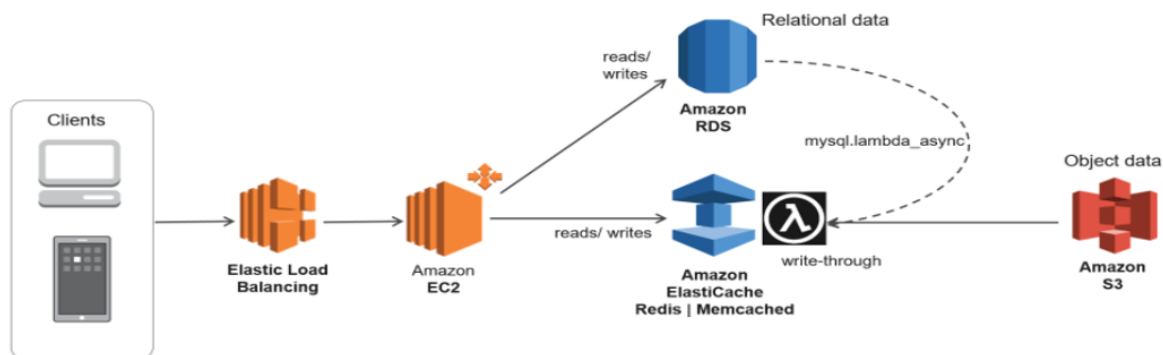

Fig6: Real-time apps and caching

**Amazon Redshift:**

- Fast, fully managed data warehouse. It is simple and cost-efficient to evaluate all data with the help of standard SQL and current BI Tools.
- Allows to run complex analytic queries against petabytes of structured data, using sophisticated query optimization, and massively parallel query execution. Most results come back in seconds.
- Delivers fast query performance by using columnar storage technology to improve I/O efficiency and by parallelizing queries across multiple nodes.
- Inexpensive. You only pay for what you use.
- Extensible. Redshift Spectrum allows you to execute queries against exabytes of data in Amazon S3 as simply as you execute queries against petabytes of data collected on local disks in Amazon Redshift, utilizing the same SQL syntax and BI tools.
- Simple. Easily automates most of the regular administrative tasks to manage, monitor, and scale your data warehouse.
- Scalable. Easily resizes the cluster up and down as the performance and capacity demands.
- Security is built-in. It encrypts data at rest and in transit with the help of hardware-accelerated AES-256 and SSL.
- Compatible. Amazon Redshift offers standard SQL and provides custom JDBC and ODBC drivers that you can download from the console.

**Amazon Neptune:**

- Fast, reliable, fully-managed graph database service makes it easy to build and run applications with highly connected datasets.
- The core of Amazon Neptune is a purpose-built. A Graph database is a High-performance optimized engine used for collecting billions of interactions and querying the graph with milliseconds latency.
- Highly available, with reading replicas, point-in-time recovery, continuous backup to Amazon S3, and replication across Availability Zones.
- Secure, with support for encryption at rest and in transit.
- Supports open graph APIs for both Gremlin and SPARQL and provides high performance for both of these graph models and their query languages.
- Highly available, durable, and ACID (Atomicity, Consistency, Isolation, Durability) compliant.
- FULLY MANAGED. You don't need to worry about database management tasks such as hardware provisioning, software patching, setup, configuration, or backups.

**AWS Database Migration Service (DMS):**

- Helps migrate databases to AWS quickly and securely. The source database continues completely in operation for the duration of the migration.

- It can migrate your data to and from the most widely used commercial and open-source databases.
- The service supports homogenous migrations such as Oracle to Oracle, as well as heterogeneous migrations between different database platforms, such as Oracle to Amazon Aurora or Microsoft SQL Server to MySQL
- Can be used for continuous data replication with high-availability.
- Simple to use. Does not require to install any drivers or applications. It does not involve changes to the source database in most cases.
- Minimal Downtime. Helps migrate your databases to AWS with virtually no downtime.
- Low Cost. Migrating a TB-size database takes up to $3. It works for both homogeneous and heterogeneous migrations for any supported databases.
- Fast and Easy to Set-up. Setting up a migration task takes just a minute in the AWS Management Console. A migration task is where you define the parameters the AWS Database Migration Service uses to execute the migration.
- Reliable. It is highly resilient and self–healing. It constantly monitors source and target databases, network, and replication.

**AWS Cloud Design Strategy:** Another benefit of AWS is while designing Cloud, the below are checked carefully.

- Downtime

- Security / Privacy

- Limited control

- Backup protection.

- All other parameters based on your business needs

**Auto-scaling and Automatic maintenance:** Automatic process is most challenging as most of the organizations are trying to migrate the process into the automation world to perform the tasks with minimum human assistance. Automation in product evolution, growth, expansion, delivery, and administration makes the AWS cloud more robust in today's World. That is one benefit of AWS cloud why many companies are trying to apply could into their business processes. Auto-scaling and auto-maintenance are two major factors why companies want to move the application to the cloud which also saves money and time. Auto scheduling of jobs, automatic backup of the database, auto deployment can be done using cloud which reduces manual tasks incredibly. AWS auto-scaling automatically monitors and regulates compute resources to uphold the maintenance of the applications hosted in the cloud. AWS cloud differs from another cloud platform as in other cloud auto-scaling can be done only for individual services. Using AWS auto-scaling options traffic can be controlled increasing the bandwidth, for example, if in a certain time the traffic goes high, then it will scale up the bandwidth and it goes automatically low once the traffic is low.

When the cloud is hosted in a web server and during the day time the customer uses the service more, the cloud will automatically scale up the bandwidth which reduces the cost. Another example is during Black Friday most of the sites take a longer time to open and place the order. Sometimes after saving the items in a cart and while paying the bill, the site goes down, once opened again the items in the curt disappears. The recommendation is to use the cloud server and enable auto-scaling options during that time when there will be huge traffic expected. This will make the customers place the order, so a company will gain, also scale down once the deals are over which will allow saving time and money for both parties, the buyer and the organization. Presently users need the fastest availability, accessibility of data and information. If we login to any social media sites, we want to view the information, pictures as soon as we log in. Delay viewing the pages will create issues and users will not use the site often. Small size companies are facing the biggest challenge to incomes based on demand. AWS cloud is the advance platform which solves all these problems.

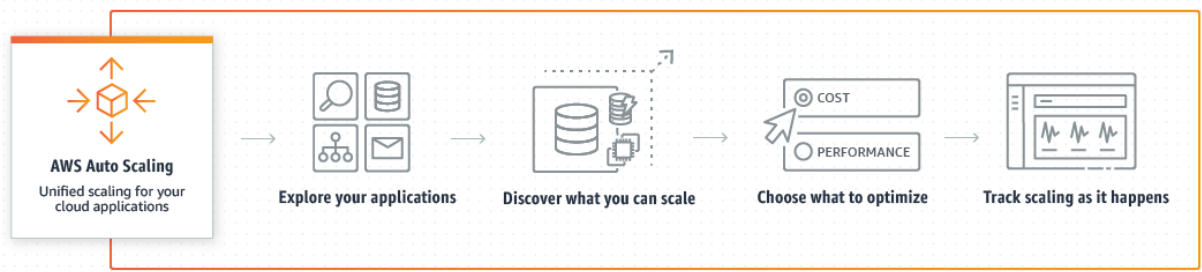

Fig7: Application scaling to optimize performance and costs [20]

**Decrease Disaster Recovery Downtime:** Zero downtime is rare during the deployment of any release. It brings the business in continuity as the system will be unavailable for a period until the deployment completes. In most of the cases, AMS cloud brings the business continuity with the disaster recovery plan that uses the cloud for storing backups and eliminate the delayed work that comes with downtime. AWS cloud platform provides much-advanced uptime rates than the normal local server. Using a multi-cloud method with backups and offsite data stored on multiple clouds can decrease the downtime to seconds or minutes hence it enables productivity.

System downtime, difficulty in failover and disaster recovery (DR) prices are major factors to have compact tools and facilities in place. However, the cumulative acceptance of AWS cloud platform, IT has added a new set of contests for protecting business crucial data that can spread outside the datacenter.

Cloud brings inclusive, money-making recovery

- Self-discovery, easy backup, universal administration of virtual machine recovery environment.
- Effective virtual machine backups are reserved at-the-ready for bringing a recovery time objective in minutes.

- Minimized recovery point objectives with global dedupe and ever-incremental backup, which also decreases storage and bandwidth expressively.
- Streamlining formation, eliminating safety and security anxieties and facilitation system failover
- No hardware or software hassle, auto- deployment

The benefits of disaster recovery are listed below –

- Fast performance: Quick retrieval of files and fast disk-based storage
- Elasticity: Fast addition of a large amount of data
- No Tape: Remove prices related with moving, loading, and recovering tape media and linked tape backup software.
- Secure: Secure and robust cloud disaster recovery platform with industry-standard guaranteed and audits.
- Compliance: Quick recovery of records permits to avoid penalties for misplaced compliance limits.
- Partners: AWS resolution provider and system incorporation partners to help with the deployments.

The following diagram shows data backup possibilities to Amazon S3, from either on-site infrastructure or from AWS

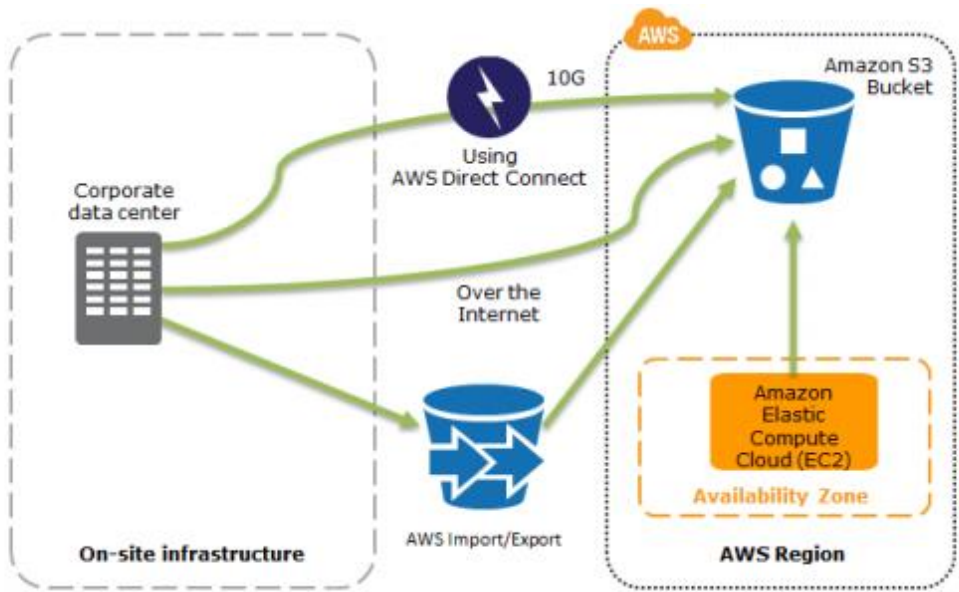

Fig8: Data Backup Options to Amazon S3 from On-Site Infrastructure or from AWS [21]

When data backup is important, more important is to restore the data for any disaster recovery. The below diagram shows how quickly the data can be restored from Amazon S3 backups to Amazon EC2.

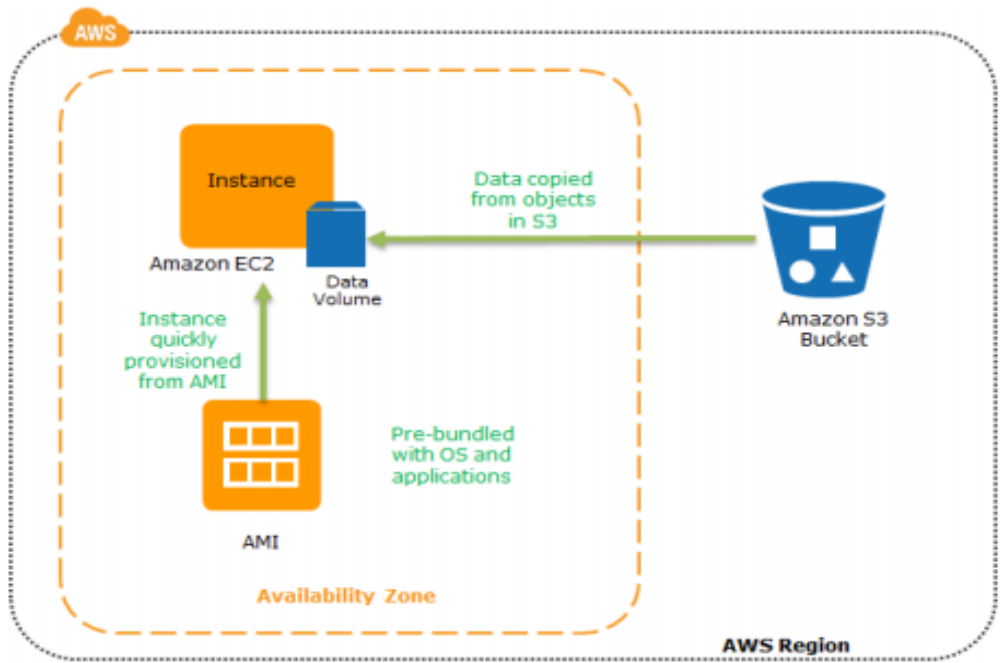

Fig9: Restoring a System from Amazon S3 Backups to Amazon EC2 [21]

**Getting edge over the competition:** AWS cloud is an advanced technique and more successful in a competitive situation. Many entities, executives, computer scientist, apps designers, corporations and industrialists use cloud computing in this competitive world. The available resources can be used to speed up the process which helps to build a good relationship with the customer and help to expand the business. Enchanting inventiveness in the execution of cloud computing in business gets ached from competition and opponents. That is one of the reasons why the cloud is essential to integrate into the project.

The facilities are trembling up the computing world in a similar technique that Amazon is varying in the retail industry. By assessing its cloud very less expensive, Amazon can offer reasonable and scalable facilities to everybody from the latest start-up to a Fortune 500 company.

Also, as we mentioned earlier the flexibility and scalability are the key reasons why AWS cloud is different from any other cloud platform. In AWS cloud the collections of objects can work together and distinctly.

There are so numerous different services are listed below -

- Compute
- Storage

- Database
- Migration
- Machine Learning
- Media Service
- Management Tools
- Game Development and even more

In AWS cloud you pay which service you choose. In another cloud platform, whether the services are used or not, payment must be done for all. That is the advance of the AWS cloud and make this different from any other providers.

## AUTHOR'S PROFILE

Sourav Mukherjee is a Senior Database Administrator and Data Architect based out of Chicago. He has more than 12 years of experience working with Microsoft SQL Server Database Platform. His work focusses in Microsoft SQL Server started with SQL Server 2000. Being a consultant architect, he has worked with different Chicago based clients. He has helped many companies in designing and maintaining their high availability solutions, developing and designing appropriate security models and providing query tuning guidelines to improve the overall SQL Server health, performance and simplifying the automation needs. He is passionate about SQL Server Database and the related community and contributing to articles in different SQL Server Public sites and Forums helping the community members. He holds a bachelor's degree in Computer Science & Engineering followed by a master's degree in Project Management. Currently pursuing Ph.D. In Information Technology from the University of the Cumberlands. His areas of research interest include RDBMS, distributed database, Cloud Security, AI and Machine Learning. He is an MCT (Microsoft Certified Trainer) since 2017 and holds other premier certifications such as MCP, MCTS, MCDBA, MCITP, TOGAF, Prince2, Certified Scrum Master and ITIL.